\documentclass[preprint,aps]{revtex4}
\usepackage{color}
\usepackage{mathtools}
\usepackage{diagbox}
\usepackage{appendix}
\usepackage{amsmath,amssymb,amsfonts,dcolumn,color,graphicx,graphics,latexsym,epsfig}
\usepackage[compatibility=false]{caption}
\usepackage{subcaption}
\usepackage{hyperref}
\usepackage{natbib}
\usepackage{float}
\usepackage{booktabs}
\def\beq{\begin{equation}}
\def\eeq{\end{equation}}
\def\barr{\begin{array}}
\def\earr{\end{array}}

\begin{document}
\title{Wormholes with a warped extra dimension?}

\author{
Sayan Kar}
\email{sayan@phy.iitkgp.ac.in}
\affiliation{Department of Physics, Indian Institute of Technology
Kharagpur, 721 302, India}

\begin{abstract}
\noindent  We investigate the role of a specifically warped extra dimension in 
constructing  examples of higher dimensional spacetimes representing Lorentzian wormholes. The warping chosen is largely inspired by the well-known non-static Witten bubble of nothing, 
though our spacetimes are static and geometrically different. Vacuum solutions in $D\geq 5$ dimensions
and others (non-asymptotically flat) with `perfectly normal' matter stress energy
are interpreted as possible Lorentzian wormholes.
Asymptotically flat wormholes in $D\geq 5$ with `exotic matter' and within this class of
spacetimes also appear to exist in all dimensions. A wormhole-black hole correspondence via double Wick rotation is
revisited and discussed. Finally, geodesic motion as well as the behaviour of geodesic congruences, in the sub-class of five dimensional, warped,
vacuum wormhole spacetimes is also briefly analysed, with the aim of obtaining
characteristic properties and specific signatures which may help improve our 
understanding of these geometries. 
\end{abstract}

\maketitle

\section{\bf Introduction} 

\noindent The idea of the `wormhole' has its precursors in the work of Flamm
\cite{flamm} on geometry of Schwarzschild spacetime (Flamm's paraboloid) and, subsequently, in the
construction of the Einstein--Rosen bridge \cite{erb} as a `particle model' in
General Relativity (GR).
Later, in the late 1950s,  Wheeler coined the term `wormhole' 
while building a topological model of 
electric charge and working on spacetimes which he christened as geons \cite{wheeler}. The Bronnikov--Ellis (BE) \cite{be,be2}
spacetime of the early 1970s is the first known wormhole solution in General Relativity (GR) {\em albeit} with a somewhat absurd, negative kinetic energy scalar field. Nevertheless, the BE spacetime is indeed a `solution'. In the 1980s and 1990s, Euclidean and Lorentzian wormholes emerged in two different contexts, the former heralded through the seminal 1988 work of Giddings and Strominger \cite{gs} and the latter
pioneered by Morris and Thorne \cite{mt} (see also \cite{mty}), also in 1988.
Since then, wormhole physics has grown into an industry with 
newer contributions on various issues and questions appearing regularly in the
literature (for  a recent review from a different perspective, see \cite{kundu}).

\noindent Let us briefly recall the definition of a Lorentzian wormhole
{\em a la} Morris--Thorne \cite{mt}. Assuming a
general static, spherically symmetric line element in four dimensions given as
\begin{eqnarray}
ds^2 = -e^{2\psi(r)} dt^2 + \frac{dr^2}{1-\frac{b(r)}{r}}
+r^2 d\Omega_2^2
\end{eqnarray}
we say that it represents a wormhole if (a)
$e^{2\psi(r)}$ has no zeros, i.e. there are no event horizons,
(b) $\frac{b(r)}{r} \leq 1$ and $b(r=b_0) = b_0$, i.e. $b_0\leq r \leq \infty$ and
(c) $\frac{b(r)}{r}$, $\psi(r)$ tend to zero  as $r\rightarrow \infty$.
The condition $(c)$ guarantees asymptotic flatness 
while the functional nature of $b(r)$ (obeying the constraints) 
gives the wormhole shape.
Thus, Lorentzian wormholes are horizon-less, 
asymptotically flat, non-singular spacetimes with their spacelike sections
having the shape representing two asymptotically flat regions connected by a bridge.
The smallest value of $r$ denoted as $b_0$ is named the wormhole throat. 
The expansion of a geodesic congruence exhibits a defocusing feature as
one crosses the throat from one side towards the other asymptotically flat region--
this being the basic reason behind the violation of the convergence condition,
as envisaged from the Raychaudhuri equation \cite{ec}.

\noindent It is thus a known fact that such static Lorentzian wormhole spacetimes \cite{mt}
cannot exist as solutions in GR with the required matter
satisfying the so-called convergence conditions (or, equivalently, in GR, the energy conditions,
such as the Null Energy Condition (NEC),
the Weak Energy
Condition (WEC)) or their averaged versions \cite{ec, visserbook} ). Therefore,
violating energy conditions is a necessity (for potential counterexamples, see
(i) \cite{bk} where a
different class of metrics is used and (ii) \cite{radu},\cite{konoplya} for 
scenarios in Einstein-Maxwell-Dirac theory). To address such violations
in a constructive sense one either tries to
justify it (eg. quantum field theoretic, effective quantum matter stress-energy or arbitrarily small `amount' of energy condition violations) \cite{qft,vkd,kdv} 
or move away from GR \cite{mg, mg1, mg2, mg3,mg4}.
Numerous examples exist in all these approaches. One can indeed have wormholes
in modified gravity without violating the energy conditions, 
as suggested in many articles in the more recent past \cite{mgrecent1, mgrecent2}.

\noindent Another approach to resolve this problem is to introduce/accept the existence of extra dimensions.
There have been earlier attempts along these lines \cite{zanganeh},\cite{wed}, \cite{bronnikov},\cite{kuhfittig}. Here, we try a different route which we now elaborate on below.

\noindent Let us consider a line element of the form:
\begin{equation}
    ds^2 = -e^{2\psi(r)}dt^2 +\frac{dr^2}{1-\frac{b(r)}{r}} +r^2 d{\Omega_2^2}
    + f^2(r) d\chi^2
\end{equation}
where $f(r)$ is the so--called warping function and $\chi$ is the extra dimension which we choose to be angular ($0\leq \chi\leq 2 \pi$).
The fact that $f$ is a function of $r$ is the reason behind the use of the term `warped'.
In recent times, the notion of warping has been heavily used in the context of braneworld
models \cite{rs}. There, the idea of a {\em warped braneworld} implies the dependence 
of the four dimensional timelike section of the higher dimensional line element, 
on the extra dimensional coordinate, usually through a conformal factor. In our line element,
as stated above, it is the extra dimensional part of the line element which is assumed
to be dependent on one (here $r$) of the so-called `four' dimensional coordinates.  

\noindent The line element is static and any $\chi=$constant section is spherically symmetric. There are Killing vectors corresponding to the coordinates $t$, $\phi$ and
$\chi$ and hence, corresponding constants of motion/conserved quantities.
Topologically the manifold is $R^2\times S^2 \times S^1$. 
In other words, $t=$constant, $r=$constant sections are toroidal -- somewhat reminiscent of the `ringholes' introduced in \cite{gonzalez} (see also \cite{bronnikov}, \cite{kuhfittig}).
We may further generalise the above line element by
introducing a $(D-3)$ sphere instead of a 2-sphere to define a $D$ dimensional line element.

\noindent A known, non-static vacuum solution in five dimensions is the
Witten bubble line element \cite{witten} given by
\begin{equation}
  ds^2 = -\alpha^2 r^2 dt^2 +\frac{dr^2}{1-\frac{b_0^2}{r^2}}
  +r^2 \cosh^2 \alpha t \, \left ( d{\theta^2 + \sin^2\theta d\phi^2}\right )
    + R^2 \left (1-\frac{b_0^2}{r^2} \right ) d\chi^2
\end{equation}
This line element may be obtained  through a double Wick rotation 
($\tau \rightarrow i \, R \chi$, $ \eta\rightarrow i \alpha t +\frac{\pi}{2}$) of the five dimensional Schwarzschild solution given by
\begin{eqnarray}
ds^2 = - \left ( 1-\frac{b_0^2}{r^2}\right ) d\tau^2 +\frac{dr^2}{1-\frac{b_0^2}{r^2}} +
r^2 \left (d\eta^2 +\sin^2\eta \, d\theta^2 +\sin^2 \eta \sin^2\theta d\phi^2\right )
\end{eqnarray}

\noindent A non-vacuum generalisation of the Witten bubble (not obtainable through any Wick rotation) discussed recently in \cite{sk21} is given as, 
\begin{equation}
  ds^2 = -\alpha^2 r^2 dt^2 + \frac{dr^2}{1-\frac{b_0^2}{r^2}}
  +r^2 \cosh^2 \rho_1 t\,d{\Omega_2^2}
    + R^2 \left (1-\frac{b_0^2}{r^2} \right ) d\chi^2
\end{equation}
where $\alpha \neq \rho_1$. If we now set $\rho_1=0$, we 
obtain a static spacetime which can be generalised further by replacing the $\frac{b_0^2}{r^2}$ in $g_{rr}$ and
$g_{\chi\chi}$ by  $\frac{b(r)}{r}$. We maintain the feature $g_{rr} g_{\chi\chi} =R^2$.
In other words, the extra dimension disappears near $r=b_0$ (the throat of the wormhole),
and is maximal as $r\rightarrow \infty$. One can also replace the  $-\alpha^2 r^2$ in $g_{00}$ with a
generic $g_{00} = -e^{2\psi(r)}$. Later, in the next section, we will work with $\psi=0$. 

\noindent In summary, what we take from the Witten bubble is the fact that the extra dimension decays away
from the asymptotic regions as we move towards the throat. As we shall see, this feature and the specific form of the line element we use helps us in ensuring that the energy conditions hold within the tenets of higher dimensional
General Relativity. As mentioned before, our wormhole has $t,r=$constant ($r>b_0$)
sections which are toroidal (i.e $S^2 \times S^1$ or, more generally, $S^{D-3} \times S^1$) with both the radii (of $S^2$ or $S^{D-3}$ and $S^1$)
increasing as we move away from $r=b_0$ towards larger values.
The $S^2$ or $S^{D-3}$ radius is ever-increasing while that of the
extra dimensional
$S^1$ is zero at the throat $r=b_0$, increases for $r>b_0$ but saturates to a finite constant $R$ as we approach $r\rightarrow \infty$ (see Figure 1).  We will now see how the specific form of the warped extra dimension influences the higher dimensional Einstein equations and leads to 
specific solutions
with distinct features.

\noindent The rest of this article is organised as follows. In Section II
we present the vacuum solutions. Section III discusses the non-vacuum case.
Finally, in Section IV geodesics and geodesic congruences are analysed.  Our concluding remarks 
appear in Section V.

\begin{figure}[ht]
\includegraphics[width=4in]{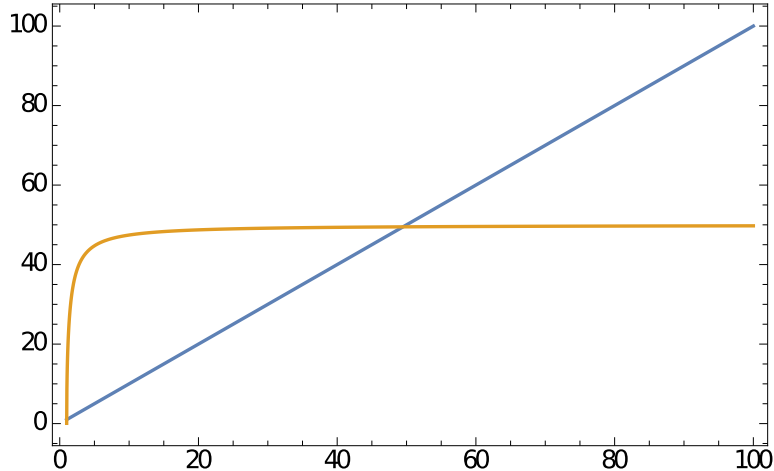}
\caption{Plot of the radii of the $S^{D-3}$ (blue) and $S^1$ (yellow)
as a function of the coordinate $r$. Here we have chosen $D=5$, $b(r)=b_0$, $b_0=1$ and $R=50$ for the line element in (6).}
\end{figure}

\section {\bf The vacuum wormhole spacetimes in $D \geq 5$}

\noindent Following our approach as mentioned in the Introduction,  we propose to work with an ultrastatic line element given as:
\begin{equation}
    ds^2 = -dt^2 + \frac{dr^2}{1-\frac{b(r)}{r}} + r^2 d\Omega_{D-3}^2 +R^2 \left (1-\frac{b(r)}{r} \right ) d\chi^2
\end{equation}
The above line element is in $D$ dimensions. $b(r)$ satisfies all requirements typical of a wormhole
but is left unspecified for the
time being. The previously (in the Introduction) mentioned $\psi(r)$ is set to zero in order to
make the geometry ultrastatic. In $D=5$, 
one can imagine the line element in Eqn. (6) as a four dimensional spacetime with a
warped extra dimension ($\chi$). When $D>5$, we adopt the viewpoint of a $D-1$ dimensional
spacetime with an extra dimension (the $\chi$). Alternatively, one may choose to
think of a $D>5$ scenario as a four dimensional spacetime with $D-4$ extra dimensions.

\noindent A curious feature of the $r=b_0$ section of the geometry is worth mentioning.
The induced four dimensional metric on this section has a determinant equal to zero--hence it is degenerate, but due to the vanishing extra dimension \cite{sandipan}.
However, it is the spatial part (not  $g_{tt}$) of the induced metric
which leads to its  degenerate
character. One therefore cannot call $r=b_0$ an event horizon (it is not an infinite redshift surface) and there is an ambiguity in referring to it as
a null surface according to the standard definition \cite{vollick}. Nevertheless, to indicate its special character, we will
refer to $r=b_0$ as a {\em degenerate throat} and discuss its characteristics
later. 

\noindent To make further progress, we need to write down the
Einstein tensors $G_{ij}$ for the metric in Eqn. (6). Thereafter, assuming $D$ dimensional
GR (i.e. $G_{ij} = \kappa T_{ij}$, $\kappa$ related to the $D$ dimensional gravitational constant) we
obtain the components of the energy--momentum tensor $T_{00}=\rho$, $T_{11} = \tau$, $T_{jj} = p$ ($j =2,3...D-2$);
$T_{D-1\,D-1} = p_{D-1}$, in the frame basis as,
\begin{eqnarray}
\rho= \frac{1}{\kappa} G_{00} = \frac{1}{\kappa} \left [  \frac{b''}{2r}
+(D-4) \left ( \frac{2 b' r +(D-5) b}{2r^3}
\right ) \right ],
\\
\tau=p_{D-1} = \frac{1}{\kappa} G_{11} = \frac{1}{\kappa}
G_{D-1 \, D-1} = 
- (D-3) \left (\frac{b'r +(D-5)b}{2r^3} \right ), \\
p=\frac{1}{\kappa} G_{jj} = \frac{1}{\kappa} \left [-\frac{b''}{2r}
-(D-5) \left ( \frac{(D-6)b+ 2 b'r}{2r^3}
\right ) \right ],
\end{eqnarray}
where, a prime denotes differentiation w.r.t. $r$.
\noindent The Ricci scalar is given as:
\begin{eqnarray}
R = \frac{b''}{r} + (D-4)\left (\frac{2b'}{r^2} + (D-5) \frac{b}{r^3} \right )
\end{eqnarray}
and the Kretschmann scalar is:
\begin{eqnarray}
K = 2(D-3) \left [ \left(\frac{b'r-b}{r^3}\right )^2 +(D-4)\frac{b^2}{r^6} \right  ] + \frac{\left (b''r^2-2 b'r+ 2b\right )^2}{r^6}
\end{eqnarray}
\noindent It is easy to see that
a vacuum solution (i.e. $G_{ij}=0$) is given by
\begin{equation}
b(r) = \frac{b_0^{D-4}}{r^{D-5}}
\end{equation}
One can arrive at the above solution by just solving the $\tau=p_{D-1}=0$
equation--the other two equations ($\rho=0$ and $p=0$) are automatically
satisfied by the solution from $\tau=p_{D-1}=0$.

\noindent Thus, for $D=5$ we have a line element given as:
\begin{equation}
    ds^2 = -dt^2 + \frac{dr^2}{1-\frac{b_0}{r}} + r^2 d\Omega_{2}^2 +R^2 \left (1-\frac{b_0}{r} \right ) d\chi^2
\end{equation}
where the $\chi=constant$ section is simply the ultrastatic spatial
Schwarzschild wormhole. For this solution, $R=0$ and $K=12 \frac{b_0^2}{r^6}$.

\noindent It must be stated that this five dimensional vacuum line element 
was, as far as our knowledge goes, first mentioned in an unpublished preprint by Roberts \cite{roberts}. 
Earlier work by Stotyn, Mann \cite{stotyn} and Miyamoto, Kudoh 
\cite{kudoh} on Einstein--Maxwell
as well as Einstein-p-form theories were concerned with related non-vacuum solutions.
More recently, Bah and Heidmann \cite{bah1}, \cite{bah2} 
have explicitly re-mentioned these types of vacuum solutions in five 
dimensions. However, the notion that these solutions could actually
represent Lorentzian wormholes has never been analysed in any detail.
Further, consequences for {\em any} $b(r)$ {\em vis-a-vis} the energy conditions,
non-asymptotically flat scenarios, geodesic motion and 
the behaviour of geodesic congruences have not been dealt with before. Our purpose in this article is to work on these aspects, to some extent.

\noindent To provide a $D>5$  example let us now
write down the line element in $D=6$. We have,
\begin{equation}
    ds^2 = -dt^2 + \frac{dr^2}{1-\frac{b_0^2}{r^2}} + r^2 d\Omega_{3}^2 +R^2 \left (1-\frac{b_0^2}{r^2} \right ) d\chi^2
\end{equation}
which, for $\chi=constant$, is a Bronnikov-Ellis like spacetime \cite{be,be2} but with
a $S^3$ instead of a $S^2$  and a warped extra dimension.  Here
$R=0$ and $K= 72\frac{b_0^4}{r^8}$. One may construct likewise, examples
in other higher dimensions too. Thus, one may state that 
this entire family of vacuum spacetimes
represent Lorentzian wormholes
in $D\geq 5$. The solution in $D=4$ (to be considered again later) 
turns out to be flat spacetime and is not obtainable directly from
the general expression provided above for $b(r)$. 

\noindent Let us then switch to an important issue--the NEC inequalities \cite{ec} for the matter required to support generic spacetimes with
any $b(r)$ satisfying the wormhole criteria.
The NEC, for our case, is stated in terms of the two expressions
for $\rho+\tau$ and $\rho+p$ (note that $\tau=p_{D-1}$, so $\rho+p_{D-1}$
is not different from $\rho+\tau$).
We have,
\begin{eqnarray}
\rho+\tau = \frac{1}{\kappa} \left [ \frac{b''}{2r} +\frac{(D-5)(b'r-b)}{2r^3} \right ] \geq 0 \\
\rho + p = \frac{1}{\kappa} \left [ \frac{b'r+(D-5) b}{r^3} \right ] \geq 0
\end{eqnarray}

\noindent To proceed further, we recall the central result emerging from the Morris-Thorne
theorem on wormhole existence and energy conditions \cite{mt}, in four dimensions.
Consider the four dimensional static, spherically symmetric line element stated
before in Eqn. (1).
A $t=$constant, $\theta=\frac{\pi}{2}$ two dimensional
section of this four dimensional line element has an induced metric given as
\begin{eqnarray}
ds^2 = \frac{dr^2}{1-\frac{b(r)}{r}} + r^2 d\phi^2
\end{eqnarray}
Embedding this section in three dimensional Euclidean space (cylindrical coordinates) with the line element
\begin{eqnarray}
ds^2 = dz^2+ dr^2 + r^2 d\phi^2
\end{eqnarray}
and defining a profile function $z(r)$ we find, from a comparison of the two dimensional line elements (Eqn. (17) and Eqn. (18) with $z=z(r)$),
\begin{eqnarray}
\frac{dz}{dr}
= \pm \sqrt{\frac{b}{r-b}}
\end{eqnarray}
The
requirement on $z(r)$ for a wormhole shape implies that $r(z)$ has a minimum at $z=0$
which corresponds to the
smallest value of $r$, i.e. $r=b_0$. The minimum ($\frac{d^2r}{dz^2}(z=0)>0$) is possible
only if $b-b'r >0$, a result which follows from the expression
for $\frac{d^2r}{dz^2}$, i.e.
\begin{eqnarray}
\frac{d^2 r}{dz^2} = \frac{b-b'r}{2 b^2}.
\end{eqnarray}
In contrast, the four dimensional Einstein equations
for the metric in Eqn (1)
imply, via the Einstein tensor and the energy conditions (assuming the Einstein equation in GR), the NEC relation $\rho+\tau= \frac{1}{\kappa} \left (\frac{b'r-b}{r^3}\right ) \geq 0$ (when $\psi=0$).
Hence, we end up with a contradiction which necessitates
the violation of the NEC if wormholes have to exist in four dimensional GR
with the added assumption that NEC must hold good.

\noindent It is easy to see that for the line elements in Eqn. (2) or Eqn. (6), the above analysis (resulting in $b'r-b<0$), from the embedding side of the argument, 
remains unaltered. This happens because the line elements in Eqn. (2)
or Eqn. (6) (for say, $D=5$) have $t=$constant, $\theta=\frac{\pi}{2}$ and $\chi=$constant two dimensional sections which are the same as the $t=$constant, $\theta=\frac{\pi}{2}$ two dimensional sections of the line element in Eqn. (1). 

\noindent Interestingly, the new expressions for 
the energy condition inequalities (Eqns. (15) and (16)) in the $D$ dimensional spacetimes, do not imply
any specific requirement on $b'r-b$ for $D\geq 5$. In fact, Eqns. (15) and (16)
do lead to restrictions on $b(r)$, but they are not directly on $b'r-b$ and 
therefore different from what is found
for the four dimensional line element in Eqn. (1). 

\noindent Thus, it is possible to have higher dimensional spacetimes 
representing wormholes and, remarkably, we
do end up with vacuum wormholes, for which there is no issue of
energy condition violation! It is notable that the warping of the
extra dimension in the manner shown in Eqn. (2) or Eqn. (6) plays a
major role in this analysis and the ensuing result.

\noindent Let us now recall another related class of
spacetimes with generic line elements of the form
\begin{eqnarray}
ds^2 = -\left (1-\frac{b(r)}{r}\right ) d\tau^2 +
\frac{dr^2}{1-\frac{b(r)}{r}}+ r^2 d\Omega_{D-3}^2 + R^2 d\xi^2
\end{eqnarray}
The line element on a $\xi=$constant slice is spherically symmetric, static and written in the Schwarzschild gauge.
The extra, unwarped compact dimension is represented by the coordinate $\xi$. The radius
of the extra-dimensional $S^1$ is $R$ and it is the same for all $r$, unlike the
earlier line element in Eqn. (2) where we had a warped extra dimension. Line elements of the type in (21) fall within the class known as black strings and branes (for a good recent
review of past literature see \cite{collingbourne}, for the specific case of vacuum solutions
see \cite{roberts}, \cite{bah1}, \cite{bah2}).

\noindent Interestingly, as is well--known \cite{bah1,bah2}, 
one can obtain from the above metric, the one given in Eqn. (6)
by a double Wick rotation -- $\tau\rightarrow i \, R \chi$
and $\xi \rightarrow \frac{i}{R} \, t$. The geometry in Eqn. (21)
could be a black hole with an extra dimension depending on the
existence of a horizon ($g_{00} (r_H) =0$) and a singularity inside the
horizon. The novelty here is quite straightforward --the horizon, topologically,  
is not just $S^{D-3}$ but $S^{D-3} \times S^1$. The Ricci scalar $R$ and
the Kretschmann scalar $K$ for this geometry are
the same as given earlier in Eqns. (10) and (11), respectively. Thus, for any $b(r)$, the geometries represented by
Eqn. (6) and Eqn (21) have the same Ricci and Kretschmann scalars. However, the Einstein tensor and
hence, the energy--momentum tensor components are different (for the non-vacuum cases) and given by:
\begin{eqnarray}
\rho^{BH}= \frac{1}{\kappa} G_{00} = -\tau^{BH}= -\frac{1}{\kappa} G_{11} = \frac{1}{\kappa}
\left [ (D-3) \left (\frac{b'r +(D-5)b}{2r^3} \right ) \right ] \\
p^{BH}= \frac{1}{\kappa} G_{jj} = \frac{1}{\kappa} \left [  -\frac{b''}{2r}
-(D-5) \left ( \frac{2 b' r +(D-6) b}{2r^3}
\right ) \right ] \\
p_{D-1}^{BH} =  \frac{1}{\kappa}
G_{D-1 \, D-1} = \frac{1}{\kappa} \left [-\frac{b''}{2r}
-(D-4) \left ( \frac{(D-5)b+ 2 b'r}{2r^3}
\right ) \right ]
\end{eqnarray}
The vacuum solution here, is once again $b(r) =\frac{b_0^{D-4}}{r^{D-5}}$ (note that
$b_0$ here can, in general, be different from the $b_0$ in the wormhole). One can
obtain this solution by solving the $\rho^{BH}=0$ equation--its solution satisfying 
the $p^{BH}=0$ and $p_{D-1}^{BH}=0$ equations automatically. The 
vacuum spacetime with the chosen $b(r)$ represents a black hole with an unwarped compact extra dimension and a toroidal horizon.
In $D=5$ it is just a Schwarzschild black hole with a $S^2\times S^1$ horizon topology. When $D=6$, we end up with a $M=0$, `$Q^2<0$' mutated Reissner--Nordstr{\"o}m spacetime (recall the Einstein-Rosen bridge! \cite{erb})
with a compact extra dimension and a similar toroidal horizon of
topology $S^3\times S^1$.
The Wick-rotated 
counterparts of all these black hole spacetimes are wormholes with toroidal 
$t,r=$ constant sections (for $r>b_0$) and a degenerate throat ($r=b_0)$. 
However, the above spacetime in (21) does have a maximal extension and can be
continued to the region $r<b_0$, i.e. inside the horizon (at $r=b_0$), for specific
choices of $b(r)$ admitting a horizon (and also a singularity inside the horizon). 
In contrast, for the wormhole obtained
by Wick rotation one must have $r \geq b_0$, otherwise, one encounters a signature change and associated pathologies. 

\noindent An important next question that may arise is -- what happens if there is matter? Is it still possible to have wormholes without
violating the energy conditions or do we need exotic matter? 
What kind of black holes (Eqn (21)) do we end up with?
We dwell on these queries briefly in the following section.

\section {\bf Non-vacuum spacetimes in diverse dimensions}

\noindent Let us first analyse the special case $D=5$. Here we find a rather unusual result for the expressions 
given  for $\rho$, $\tau$, $p$ and $p_4$.
Writing them down explicitly,  we find
\begin{eqnarray}
\rho = \frac{1}{\kappa} \left ( \frac{b''}{2r} + \frac{b'}{r^2}\right )
\\ 
\tau = p_4 = -\frac{1}{\kappa} \frac{b'}{r^2} \\
p=-\frac{1}{\kappa} \frac{b''}{2 r}
\end{eqnarray}
Notice that only derivatives of $b$ appear. 
If we now consider the WEC or the NEC, it is easy to see that they will be satisfied as long 
as $b(r)$ and its first two derivatives are always positive, i.e. $b'$ and $b''$ are always 
greater than zero. A standard example could be 
\begin{equation}
    b(r) =  b_0^\nu r^{1-\nu} \hspace{0.2in}; \hspace{0.2in} 
    b'(r) = b_0^\nu (1-\nu) r^{-\nu} \hspace{0.2in} ; \hspace{0.2in}
    b''(r) = b_0^\nu \nu (\nu-1) r^{-\nu-1}
    \end{equation}
    where $\nu \leq 0$ if $b$, $b'$ and $b''$ are to be positive.
    However, such a choice of $\nu$ does not yield an asymptotically flat
    spacetime or a wormhole. It is only when $\nu=1$, i.e. vacuum, that we
    get an asymptotically flat spacetime and a wormhole without any
    energy condition violations. On the other hand, if we allow
    violation of the energy conditions we can surely have non-vacuum
    wormholes for $0\leq \nu\leq 1$.
    
 \noindent Beyond $D=5$, a similar result persists. For general $D$ and with
 the above choice of $b(r)$, the inequalities in Eqns. (15) and (16)
 lead to the relations $\rightarrow$  $\nu (\nu-D+4) \geq 0$ and $(D-4-\nu) \geq 0$, respectively.
 For $0\leq \nu\leq 1$ both relations cannot hold simultaneously. In contrast,
 when $\nu <0$ they lead to a single inequality $(\vert \nu\vert +D-4) \geq 0$,
 which can indeed be satisfied but leads to a non-asymptotically flat
 spacetime. 
Thus, it is only the vacuum spacetimes ($\nu= D-4$) mentioned earlier which can have the features of an 
asymptotically flat Lorentzian wormhole, as long as energy conditions are to be respected with `matter' (here, vacuum) defined via the higher dimensional
Einstein field equations. This statemnent is of course restricted to the class of $b(r)$ mentioned above in Eqn. (28). For other choices of $b(r)$ it may be possible to
restrict the region over which energy condition violations occur. One can indeed play around with different choices of $b(r)$ and analyse the resulting consequences and differences with the
standard four dimensional wormholes.
   
\noindent An interesting case arises when $b(r)= \Lambda r^3$ (i.e. $\nu=-2$, 
$\Lambda= \frac{1}{b_0^2}$, $D=5$).
This yields (using the expressions for $D=5$)
 \begin{eqnarray}
 \rho = \frac{6\Lambda}{\kappa} \hspace{0.2in}; \hspace{0.2in}
 \tau = p_4 = -\frac{3\Lambda}{\kappa} \hspace{0.2in} ; \hspace{0.2in}
 p=-\frac{3\Lambda}{\kappa}
 \end{eqnarray}    
 For $\Lambda >0$, the energy conditions hold and the Ricci scalar
 $R=12 \Lambda$. If we now write a new $b(r) = \Lambda r^3 + b_1$ (
 yielding the same values on the R. H. S. in the above equations) 
 we can obtain a non-asymptotically flat Lorentzian wormhole satisfying the energy conditions with its throat radius
 given by $b_0= \sqrt{\frac{\beta-1}{\beta}} \frac{1}{\sqrt{\Lambda}}$ ($\beta >1$) where
 $b_0=\beta b_1$. Hence, if we give up asymptotic flatness then there is a possibility of constructing wormholes with normal matter. 
 
\noindent Another important question is -- what happens for $D=4$ for this class of metrics? 
To answer this
we write down $\rho$, $\tau$ and $p$ in four dimensions.
\begin{eqnarray}
\rho = \frac{1}{\kappa} \left ( \frac{b''}{2r}\right )
\\ 
\tau = p_3= \frac{1}{\kappa} \left ( \frac{b-b'r}{2r^3} \right ) \\
p=\frac{1}{\kappa} \left ( -\frac{b''}{2 r} +\frac{b'r-b}{r^3}\right )
\end{eqnarray}
Now the $\rho+p$ inequality requires $b'r-b>0$ which contradicts the 
condition $b-b'r>0$, for a wormhole, from the embedding analysis.
Thus, in $D=4$, one ends up with NEC violating exotic matter 
for such wormholes with toroidal ($S^1\times S^1$) $r>b_0$ sections and a  degenerate
throat, to exist within the framework of GR. Unlike what we found for $D\geq 5$, in $D=4$, for the given class of metrics, giving up asymptotic flatness
cannot rescue us from avoiding energy condition violations. 

\noindent Finally, let us see how the stress-energy ($\rho^{BH}$, $\tau^{BH}$, $p^{BH}$ and $p_4^{BH}$) 
required to support a black hole (Eqn. (21)) are related to that for the wormhole
(i.e. $\rho$, $\tau$, $p$, and $p_4$). We restrict to $D=5$ and compare Eqns (22), (23), (24)
with (7), (8), (9) to get
\begin{eqnarray}
\rho^{BH} = -\tau = \frac{1}{\kappa} \frac{b'}{r^2} \hspace{0.2in};\hspace{0.2in} 
\tau^{BH} = \tau = -\frac{1}{\kappa} \frac{b'}{r^2} \\
p^{BH} = p = -\frac{1}{\kappa} \frac{b''}{2r} \hspace{0.2in}, \hspace{0.2in} 
p_4^{BH} = -\rho = \frac{1}{\kappa} \left ( - \frac{b''}{2r}-\frac{b'}{r^2} \right )
\end{eqnarray}
Thus, for the wormhole, as we noted earlier,
energy conditions will hold if $b'>0$, $b''>0$.
In contrast, for the black hole, we require $b''<0$ and $b''<\frac{2b'}{r}$.
For example when $b(r) = 2M -\frac{Q^2}{r}$ 
we obtain a Reissner-Nordstr{\"{o}}m black hole (with an extra dimension) with matter satisfying the energy conditions. With the same $b(r)$, the wormhole geometry is generated with NEC violating matter.
In general, $\rho^{BH} + \tau^{BH}=0$ but $\rho^{BH} + p^{BH} = p-\tau$ and $\rho^{BH}+p_4^{BH}= - \left (\rho+\tau \right )$. Hence we have the following intriguing result: {\em a NEC violating wormhole
($\rho+\tau<0$ but $p \geq \tau$) could correspond to a NEC satisfying black hole ($\rho^{BH} +\tau^{BH}\geq 0$
as well as $\rho^{BH}+p^{BH}, p_4^{BH} \geq 0$) for the same $b(r)$!}  

\noindent In order to further
appreciate and unravel specific characteristics of these spacetimes in $D\geq 5$, let us
now focus on particle trajectories.

\section{\bf Geodesics}
\noindent Among many possible studies which can be done in any new spacetime, knowing about geodesic motion is perhaps a first.
We will study geodesics in the five dimensional vacuum geometry where
$b(r)=b_0$. The geodesic Lagrangian is given as:
\begin{eqnarray}
{\cal L} = -{\dot t}^2 + \frac{{\dot r}^2}{1-\frac{b_0}{r}} + r^2 {\dot\theta}^2 + r^2 \sin^2 \theta {\dot \phi}^2 + R^2 \left (1-\frac{b_0}{r}\right ) {\dot \chi}^2 = -k
\end{eqnarray}
where $k=1$ (timelike) or $k=0$ (null).
From the first integrals for $t$, $\phi$ and $\chi$ we obtain,
\begin{eqnarray}
\dot  t  = E \hspace{0.2in}; \hspace{0.2in} \dot \phi = \frac{L}{r^2}
\hspace{0.2in} ; \hspace{0.2in} \dot \chi = \frac{C}{R^2 \left (1-\frac{b_0}{r}\right )} 
\end{eqnarray}
where we have chosen $\theta = \frac{\pi}{2}$ without any loss of generality. $E$, $C$ and $L$ are the constants of motion. In particular,
$C$ is associated with the extra dimension $\chi$.

\noindent Using the above expressions, it is easy to find $\dot r$, which is given as,
\begin{eqnarray}
\dot r = \pm \sqrt{\left (E^2-k-\frac{L^2}{r^2} \right ) \left (1-\frac{b_0}{r}\right ) - \frac{C^2}{R^2}}
\end{eqnarray}
and hence, an effective `potential' is
\begin{eqnarray}
V_{eff} (r) = - \frac{1}{2} \left (E^2-k-\frac{L^2}{r^2} \right ) \left (1-\frac{b_0}{r}\right ) + \frac{C^2}{2 R^2}
\end{eqnarray}
Note that ${\dot r}^2 = -V_{eff}(r)$ and hence, $r$ values where $V_{eff} \leq 0$ 
correspond to regions where physical motion is allowed.

\noindent For $L=0$ it is possible to obtain $t(r)$ in terms of simple functions
though the expression is not invertible.
We find, after integrating,
\begin{eqnarray}
t = \sqrt{\frac{r(\alpha r - \beta)}{\alpha^2}} +\frac{\beta}{\alpha^{\frac{3}{2}}} {\sinh}^{-1} \sqrt{\frac{\alpha r -\beta}{\beta}}
\end{eqnarray}
where $\alpha = 1- \frac{k}{E^2} -\frac{C^2}{E^2 R^2}$ and  
$\beta = b_0 \left (1-\frac{k}{E^2}\right )$. 
Note that $t=0$ at $r= \frac{\beta}{\alpha} <b_0$, as long as $C\neq 0$.
Thus, the presence of the extra dimension is reflected in the fact that
trajectories may not quite reach the wormhole throat unless $C=0$.
This could be thought of as a signature of the presence of the extra
dimension. Figure 1 shows the solution for $t(r)$ in the left panel
for $C\neq 0$ and $C=0$. The corresponding effective potentials are
shown on the right panel. One notices the point where $V_{eff}$ is zero and then
the region where it is negative which, together reflect the feature just mentioned above. 
\begin{figure}[ht]
\begin{minipage}[b]{0.45\linewidth}
\centering
\includegraphics[width=\textwidth]{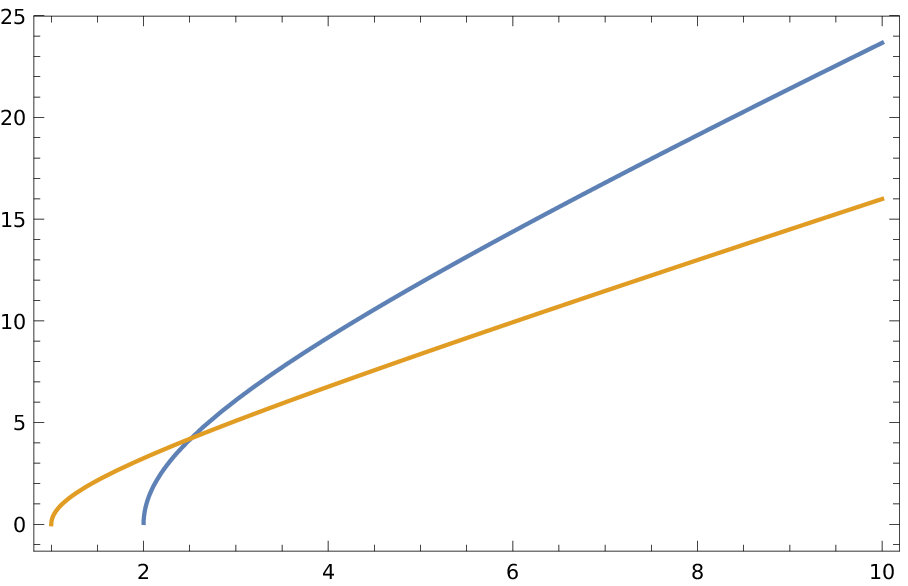}
\label{fig:figure1}
\end{minipage}
\begin{minipage}[b]{0.45\linewidth}
\centering
\includegraphics[width=\textwidth]{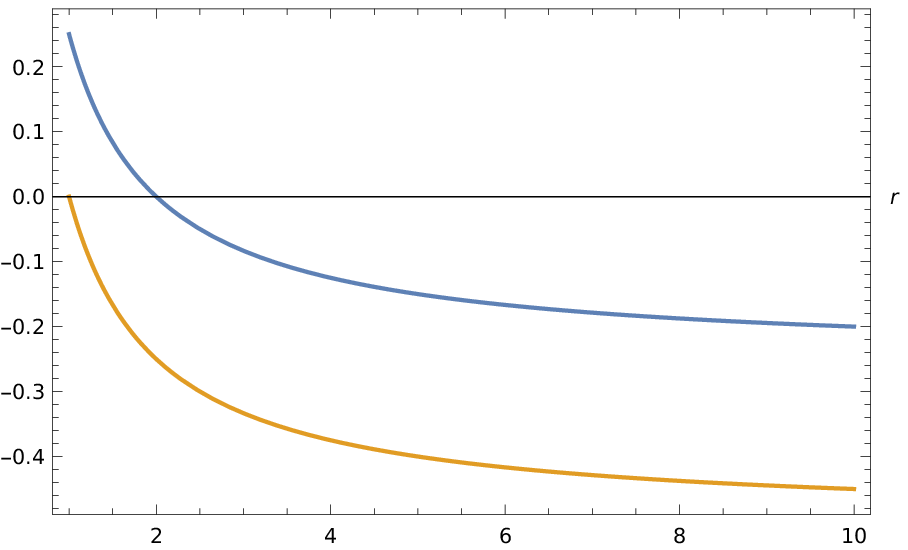}
\label{fig:figure2}
\end{minipage}
\caption{The plots refer to the solution $t(r)$ in (39) and the $V_{eff}$ in (38).
Left: $t$ ($y$-axis) vs. $r$ (x-axis) for $k=1, b_0=1, E^2=2, C^2=0.5, R^2=1, L=0$ [blue] and $k=1, b_0=1, E^2=2, C^2=0, L=0$ (yellow);  
Right: $V_{eff} (r)$  ($y$ axis) vs $r$ ($x$ axis) 
for $k=1, b_0=1, E^2=2, C^2=0.5, R^2=1, L=0$ [blue] and $k=1, b_0=1, E^2=2, C^2=0, L=0$ (yellow)}
\end{figure}

\noindent When $L\neq 0$, the solution for $t(r)$, though obtainable, is quite complicated.
In order to understand what are the consequences for $L\neq 0$, it is
useful to look at $\left(\frac{dr}{dt}\right )^2$ which we rewrite below.
\begin{eqnarray}
\left ( \frac{dr}{dt} \right )^2 = \frac{\alpha}{r^3} \left [ r^2(r- \gamma b_0)-\eta (r-b_0)\right ]
\end{eqnarray}
Here $\gamma = \frac{E^2-k}{\alpha E^2} >1$ and $\eta = \frac{L^2}{E^2 \alpha}$. When $C=0$
we have 
$\gamma=1$ and any test particle can start out from or reach $r=b_0$, as long as
$\eta \geq b_0^2$ (or $L^2 > \alpha b_0^2$). When $C\neq 0$ along with $L\neq 0$
we can see that $r=b_0$ is not reachable--a result similar to what we found in the
$L=0$ case. Hence, in our class of metrics, the inaccessibility of the throat
at $r=b_0$ seems an unavoidable signature of the presence of extra dimensions. 
The difference in the asymptotic value of the effective potential
for the $C=0$ and $C\neq 0$ cases as well as the presence of the extra dimension
in the asymptotic region are both responsible for the above-mentioned behaviour of 
test particles.

\noindent It is also possible to arrive at the above mentioned conclusion 
by looking at the expansion ($\Theta$) of a timelike geodesic congruence.
To understand this better, we rewrite the general line element in the form:
\begin{eqnarray}
ds^2 = -dt^2 + dl^2 + r^2(l) d\Omega_2^2 + R^2 {r'}^2(l) d\chi^2
\end{eqnarray}
using the proper radial distance $l(r)$ obtained from
\begin{eqnarray}
\frac{dl}{dr} = \pm \frac{1}{\sqrt{1-\frac{b_0}{r}}}
\end{eqnarray}
The inverse of $l(r)$ is $r(l)$ and $r'(l) = \frac{dr}{dl}$.
The normalised timelike geodesic vector field $u^i$ is therefore given as:
\begin{eqnarray}
u^i \equiv =\left (u^t,u^l,u^\theta, u^\phi, u^\chi \right ) \equiv \left ( E, \pm \sqrt{E^2-1-\frac{L^2}{r^2}-\frac{C^2}{R^2 {r'}^2}}, 0, \frac{L}{r^2}, \frac{C}{R^2 {r'}^2} \right )
\end{eqnarray}
where $\theta=\frac{\pi}{2}$ and $r(l)$ is left unspecified though we assume, as before, its
wormhole features.
The expansion $\Theta$ of a timelike geodesic congruence, i.e. $\Theta = \nabla_i u^i$ turns out to be:
\begin{eqnarray}
\Theta (l) = \frac{1}{u^l} \left ( \frac{r'}{r} \left \{ 2(E^2-1)-\frac{L^2}{r^2}- 2 \frac{C^2}{R^2 {r'}^2} \right \} +\frac{r''}{r'} \left \{ E^2-1-\frac{L^2}{r^2}\right \} \right ) 
\end{eqnarray}
For $L,C=0$, the expansion takes the simpler form
\begin{eqnarray}
\Theta (l) = \sqrt{E^2-1} \left ( 2\frac{r'}{r} + \frac{r''}{r'}\right )
\end{eqnarray}
where the second term is due to the warped extra dimension. One notices that
For $L,C=0$, the caustic in
$\Theta$  (locus of points where $\Theta \rightarrow -\infty$) arises at the location ($l$ value) where $r'(l)=0$, which is the location of the throat.
However, when $C\neq 0$, the caustic location shifts away from $l=0$, a fact derivable from the zero value of the denominator factor $u^l$ in Eqn. (44). In a four dimensional ultrastatic wormhole spacetime the expansion will just involve $\frac{r'}{r}$ and hence there will be defocusing near the throat and no caustic formation. Further, if instead of
the line element in (41), we write (21) using the $l$ coordinate, we have
\begin{eqnarray}
ds^2 = -{r'}^2(l) dt^2 + dl^2 + r^2(l) \left (d\theta^2 +\sin^2\theta d\phi^2\right )
+ R^2 d\xi^2
\end{eqnarray}
It can be shown that the expansion $\Theta$ of the timelike geodesic congruence
will not diverge at the horizon at $l=0$ -- in fact it will be finite and negative
for any $r(l)$ (or $b(r)$) which satisfies wormhole-like features.

\noindent One may therefore, identify the presence of the extra dimension through the
location of the caustic. For null geodesics, a parallel analysis can also be done with
qualitatively similar consequences.

\noindent On the other hand, one may obtain circular orbits by 
solving $V_{eff} (r_0)=0$ (i.e. zero radial velocity or $\frac{dr}{dt} =0$)
which gives
\begin{eqnarray}
\left (\frac{r_0-b_0}{r_0}\right ) \left (\frac{r_0^2 (E^2-k)- L^2}{r_0^2}
\right ) = \frac{C^2}{R^2}
\end{eqnarray}
Notice that $r_0=b_0$ only when $C=0$. When $C\neq 0$ one has to solve a general  cubic equation for $L\neq 0$. We note that
there are timelike and null orbits, i.e. for $k=1$ as well as 
$k=0$. The cubic equation in $\frac{1}{r_0}$ can be reduced to a depressed cubic in the variable $y$ by using the transformation
$\frac{1}{r_0} = y + \frac{1}{3b_0}$. We get
\begin{eqnarray}
y^3 + s y + q =0
\end{eqnarray}
where
\begin{eqnarray}
s = -\frac{1}{3b_0^2} - \frac{E^2-k}{L^2}  = -\frac{1}{3b_0^2} - \frac{1}{{L'}^2} \\
q= \frac{1}{L'^2 b_0} \left [ \frac{2}{3} - \frac{2 {L'}^2}{27 b_0^2} -d_0^2 \right ]
\end{eqnarray}
with ${L'}^2 = \frac{L^2}{E^2-k}$ and $d_0^2 = \frac{C^2}{R^2(E^2-k)}$.
It is easy to see that $r_0=3b_0$ (i.e. $y=0$) is a solution when
$L'^2=b_0^2$ and $d_0^2==\frac{16}{27}$. A general solution
can also be written down using the trigonometric method of finding the roots of a cubic.

\noindent When $L=0$, expectedly, the analysis is easier. The solution for
$r_0$ is simple and given as:
\begin{eqnarray}
r_0 = \frac{b_0}{1- \frac{C^2}{R^2 (E^2-k)}}
\end{eqnarray}
which equals $b_0$ for $C=0$ and is always greater than
$b_0$ as long as $C^2< R^2 (E^2-k)$. Choosing $k=1
$ or $k=0$ one can find results for timelike and null 
circular orbits. It is important to note the condition that
the circular orbit at $r_0$ is a closed curve on the
torus defined by $\phi$ and $\chi$ (recall $\theta=\frac{\pi}{2}$). 
Since we have,
\begin{eqnarray}
\dot \phi = \frac{L}{r_0^2} \hspace{0.2in} ; \hspace{0.2in}
\dot \chi =  \frac{C}{R^2 \left (1-\frac{b_0}{r_0}\right )}
\end{eqnarray}
and, therefore,
\begin{eqnarray}
\phi = \frac{L}{r_0^2} \frac{R^2 \left (1-\frac{b_0}{r_0}\right )}{C}
\chi
\end{eqnarray}
The closure of the curve is dependent on the requirement 
\begin{eqnarray}
\frac{L}{r_0^2} \frac{R^2 \left (1-\frac{b_0}{r_0}\right )}{C} = 1
\end{eqnarray}
For example, if $r_0=3b_0$ one ends up with the condition
that $\frac{R^2}{b_0^2} = \frac{27}{2}\frac{C}{L}$, which constrains the
asymptotic radius of the extra dimension in terms of the wormhole throat.
In principle, one may replace the R.H.S. in Eqn. (48) by an integer `$n$'
which will imply multiple windings along one direction as equivalent to
a single winding along the other.

\noindent In summary, one may have three types of closed curves on the `torus'
(assuming $\theta=\frac{\pi}{2}$)
with coordinates defined by
$\phi$ and $\chi$. Constants of motion related to $\phi$ and $\chi$ are  $L$ and
$C$, respectively. When $L=0$, the closed curves are defined by $\chi$ with $C\neq 0$
for $r_0 >  b_0$ (strictly). 
If $C=0$, similar closed curves are defined via $\phi$ with $L\neq 0$ and with $r_0\geq b_0$.
In general, one may have closed curves with both $L$ and $C$ nonzero for 
$r>b_0$ and with a constraint involving $L$, $C$ and $r_0$ which is required
for closure. The distinct character of the closed curves for $C\neq 0$ is a
signature of the presence of the extra dimension $\chi$.

\section{Conclusions}

\noindent To conclude, we summarise our results with comments.

\noindent Firstly, we have  looked at a class of higher dimensional warped line elements 
in which, the spacelike section has wormhole features with toroidal $r>b_0$ sections
and a degenerate throat. For example,
in $D=5$, the $r>b_0$ sections are, topologically, $S^2\times S^1$. The nature of the warping
(inspired by the Witten bubble geometry) is such that in $D=5$, the
extra dimension is maximal in the asymptotic region and has a zero radius at the throat
(degenerate throat). 
Such a higher dimensional wormhole spacetime is shown to exist in vacuum which means that
there is no issue with energy conditions or their violation. For $D>5$, the $D-1$ dimensional
timelike section represents a higher dimensional wormhole with an additional compact extra dimension. We have written down the general $D$ dimensional vacuum line element. We also show how
a black hole line element with a unwarped extra dimension (a black string) 
can be arrived at using a double Wick rotation of the wormhole metric--thereby revisiting a known correspondence \cite{bah1,bah2},
with the `wormhole' aspect added. In a broader sense, the two types
of spacetimes, i.e. wormholes and black holes seem to exist for different
sets of values of parameters in a {\em complex} line element of the form
\begin{eqnarray}
ds^2 = -\left (1-\frac{b(r)}{r} \right ) a^2 e^{2 i \alpha} d\tau^2
+\frac{dr^2}{1-\frac{b(r)}{r}} + r^2 d\Omega_{D-3}^2 + R^2 d^2 e^{2 i\beta}
d\chi^2.
\end{eqnarray}
Choosing the pairs $(a,\alpha)$ and $(d,\beta)$ one can obtain the wormhole and the black hole line elements and also write down the double Wick rotations
involved in relating the metrics. 

\noindent A further extension
of this link between black holes (black strings) and wormholes is also possible. One can 
relate wormholes of different types--for example, ultrastatic ones 
(zero gravitational redshift) with non-ultrastatic ones 
(finite gravitational redshift). Details along these directions will be
discussed in future {\cite{sk2022}}.

\noindent Switching over to similar metrics with matter, we find non-asymptotically flat spacetimes which represent a wormhole with an extra dimension without any exotic matter. We also delineate the conditions under which any black hole or wormhole (represented by our class of metrics) could exist, with energy-condition conserving/violating matter. Numerous examples are presented. 

\noindent Finally, we analyse the geodesics in the $D=5$ vacuum 
wormhole line element and show how
the presence of an extra dimension is manifest in the trajectories of 
test particles. Importantly, we have, through an analysis of the expansion of a timelike geodesic congruence,
shown that the throat is a benign caustic when the geodesics have a constant $\chi$ (extra dimensional coordinate). For a varying $\chi$ this caustic shifts to values larger than the throat radius but is still present. 

\noindent In a sense, we have found a way to evade the Morris-Thorne theorem by exploiting a warped higher dimensional
spacetime which leads to vacuum wormholes or non-asymptotically flat ones
with matter. One may argue that these wormholes are not
`true, traversable wormholes' because they admit
geodesic congruences which end at a benign caustic. In fact, the appearance of the
caustic is the precise reason why 
they exist without violating the convergence or, equivalently for GR, the energy conditions.   

\noindent An important issue related to both the wormhole and the black hole spacetimes
with warped and unwarped extra dimensions respectively, concerns perturbations and
stability. It is known that the black string is unstable--a result famously known as
the Gregory--Laflamme instability \cite{gl1}, \cite{gl2}, \cite{gl3}, \cite{collingbourne}. 
What happens for the wormholes with a warped extra dimension? 
Further, for both the wormholes and their Wick--rotated counterparts, what happens for
different choices of $b(r)$? In particular, to answer these questions one would have to
do a detailed study of scalar as well as gravitational perturbations. It is quite likely that
such studies can lead to interesting new results for both types of spacetimes for varied choices of
$b(r)$.

\noindent Going further, it is very much possible to generalise the results here
by (a) using different functions  (different  $b(r)$) in $g_{rr}$ and $g_{\chi\chi}$ (b) removing ultrastaticity by incorporating  $e^{2\psi(r)}$ in $g_{tt}$.
One may thereby expect more flexibility in constructing 
astrophysically (and perhaps observationally!) useful, 
asymptotically flat wormholes and it is likely that 
energy-condition violations for matter stress-energy may 
eventually be avoidable, with the caveat that extra/higher dimensions are indeed around!

\section*{Acknowledgements}
\noindent I thank the editors of this special volume for inviting me to contribute an article in this issue
dedicated to the memory of Professor Thanu Padmanabhan. Thanks also to Sumanta Chakraborty,
Sandipan Sengupta and Amitabh Virmani for their valuable comments and suggestions on
the manuscript.
It is indeed an honour for me to present this article as a modest tribute to the memory
of Professor Padmanabhan.

\end{document}